\documentclass[]{iopart}

\usepackage[dvips]{graphicx}

\newcommand{\ket}[1]{\vert #1\rangle}

\newcommand{\ii}{\mathrm{i}}

\begin{document}

\title{Quantum logic via optimal control in holographic dipole traps}

\author{U. Dorner$^+$, T. Calarco$^*$, P. Zoller\S}
\address{$^+$ Clarendon Laboratory, University of Oxford, Parks Road,
  Oxford OX1 3PU, United Kingdom}
\address{$^*$ ECT*, I-38050
  Villazzano (TN), Italy and CRS BEC-INFM, Universit\`a di Trento,
  I-38050 Povo (TN), Italy}
\address{\S Institute for Theoretical
  Physics, University of Innsbruck, and Institute for Quantum Optics
  and Quantum Information of the Austrian Academy of Sciences, A-6020
  Innsbruck, Austria}

\author{A. Browaeys, P. Grangier}
\address{Laboratoire Charles Fabry de
  l'Institut d'Optique, Centre Universitaire, B\^atiment 503, F-91403
  Orsay, France}

\begin{abstract}
  We propose a scheme for quantum logic with neutral atoms stored in
  an array of holographic dipole traps where the positions of the atoms
  can be rearranged by using holographic optical tweezers. In
  particular, this allows for the transport of two atoms to the same
  well where an external control field is used to perform gate
  operations via the molecular interaction between the atoms.  We show that optimal
  control techniques allow for the fast implementation of the gates with high fidelity.
\end{abstract}

\maketitle

\section{Introduction}

In the search for a suitable system for quantum information
processing, certain requirements have to be met \cite{04}, such as
scalability of the physical system, the capability of initializing
and reading out the qubits, and the possibility of having a set of
universal logic gates. Neutral atoms are one of the most promising
candidates for storing and processing quantum information. A qubit
can be encoded in the internal or motional state of an atom, and
several qubits can be entangled using atom-light interactions or
atom-atom interactions. Schemes for quantum gates for neutral
atoms have been theoretically proposed, that rely on dipole-dipole
interactions \cite{02,qg1,qg2,qg3} or controlled collisions
\cite{03,jaksch,10,12}. Such schemes can be implemented in optical
lattices with a controlled filling factor, as shown in ref.
\cite{bloch} where multi-particle entanglement via controlled
collisions was demonstrated.

Presently a major challenge 
is to combine controlled collisions with the loading and the
addressing of individually trapped atoms. Recently techniques to
confine single atoms in micron-sized \cite{07,RS,01} or larger
\cite{meschede} dipole traps have been experimentally
demonstrated.  A set of qubits can be obtained by creating an
array of such dipole traps, each one storing a single atom
\cite{register}. Gate operations require the addressability of
individual trapping sites and reconfigurability of the array.
Arrays of dipole traps, each containing many atoms, were obtained
using either arrays of micro-lenses \cite{09} or holograms
\cite{salomon}.

Actually, holographic techniques allow one to realize arrays of
very small dipole traps \cite{grier}, which can trap single atoms.
Holographic optical tweezers use a computer designed diffractive
optical element to split a single collimated beam into several
beams, which are then focused by a high numerical aperture lens
into an array of tweezers. Recently holographic optical tweezers
for individual Rubidium atoms have been implemented by using
computer-driven liquid crystal Spatial Light Modulators (SLM)
\cite{Bergamini2004}. The advantage of these systems is that the
holograms corresponding to various arrays of traps can be
designed, calculated and optimized on a computer. As a
consequence, the trap array can be (slowly) controlled and
reconfigured by writing these holograms on the SLM in real-time.

Here we want to combine such an holographic array with a fast
moving tweezer, in order to implement quantum gates based on a
state selective collision between two atoms, by using a Feschbach
resonance. Optimizing the control of the atoms motion is then of
crucial importance, and is the subject of the present paper.

\section{Quantum register with holographic dipole traps}

The present approach for neutral atoms quantum gates is related to
several schemes which have been proposed for trapped ions
\cite{wineland,zoller}, and it uses a quantum register made of
individual atoms stored in an array of holographic dipole traps.
The atoms encoding the qubit will be stored in this register,
which can be slowly reconfigured to move the atoms around, but
does not allow fast precise motion, which is required to implement
a controlled collision between two atoms. As a consequence, the
register has to be combined with one (or several) fast tweezers,
which can rapidly move an atom from one place to an other. There
are then several options~: either there is an atom in the moving
tweezer, which can be entangled and disentangled with the atoms in
the register (``moving head" scheme, similar to the one proposed
in ref. \cite{zoller}). One can also consider a configuration with
two tweezers, which catch two atoms in the register and bring them
to interaction.

The fast tweezer (or tweezers) consist of a laser beam passing
through an acousto-optical modulator (AOM), which allow to control
simultaneously the deflection and the intensity of the beam with
high accuracy. In the present paper, we will consider only two
such tweezers, each containing one atom, and we will show that a
quantum gate can be implemented with high fidelity by using
optimal control techniques. The parameters of the calculations
will be inspired by the experiment described in ref.
\cite{01,07,RS,Bergamini2004}, but the scheme may work as well in
a large range of parameter values. Typically, the size of the beam
waist for the tweezer will be less than a micron, resulting in
oscillation frequencies  of 130 kHz in the radial directions, and
about 30 kHz in the axial direction. In addition, we will assume
that a standing wave is added along the propagation axis. This has
two important consequences~: first, the axial oscillation
frequency is increased up to a value which is typically close or
above the radial oscillation frequency; second, it will confine
the two atoms within the same ``pancake", therefore maximizing the
non-linear phase shift acquired during a controlled cold
collision. In the following, we will also assume that the two
atoms have been prepared in the ground state of the tweezer.
Though this was not implemented yet, it can in principle be done,
by using either side band cooling, or evaporative cooling down to
the single atom level \cite{RS}.

\section{Atom transport in a time-dependent double-well potential}

The transport mechanism is discussed in~\cite{calarco2004} for
atoms in a time dependent, optical super lattice which has the
form of a periodic array of double well potentials.  Here,
however, we consider a one-dimensional system with a single double
well potential of the form
\begin{equation}
V(x,t) = -A(t) \; e^{-x^2/2w^2} - B(t) \; e^{-(x+d(t))^2/2w^2}.
\label{potential}
\end{equation}
\begin{figure}[h!]
\begin{center}
  \includegraphics[]{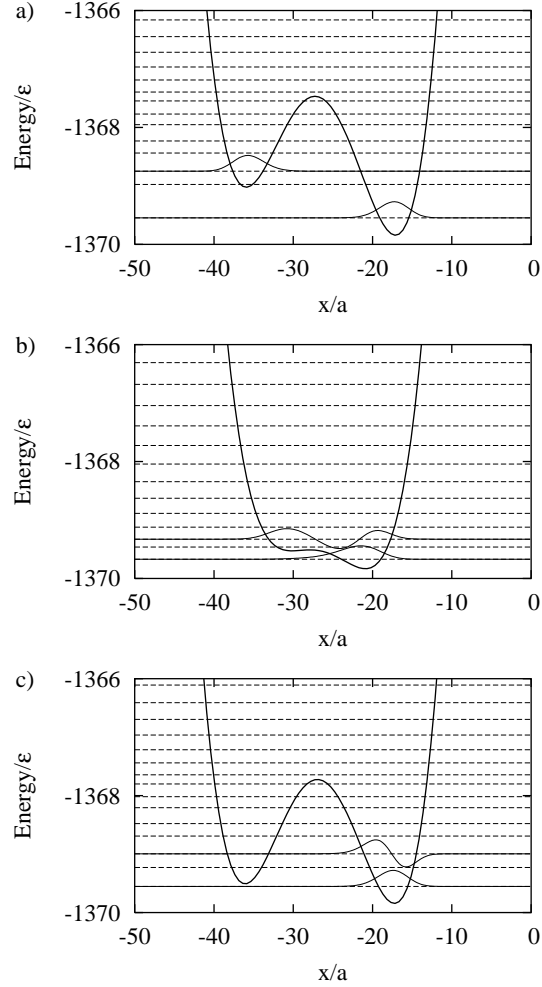}
\end{center}
\caption{
  The double well potential at (a) the initial time $t=0$, (b) at an
  intermediate time $t=T/2$ and (c) at the end of the transport
  process $t=T$. The position is given in units of
  $a\equiv\sqrt{\hbar/m\omega}$ and the energy in units of
  $\varepsilon\equiv{\hbar^2/2ma^2}$ as described in the text. The
  horizontal, dashed lines indicate the eigenenergies of the system.
  The solid lines are the (real) eigenfunctions of $H(t)$
  corresponding to the moving and the register atom, i.e.
  $\psi_2(x,t)$ and $\psi_0(x,t)$, respectively.
  The shown potentials correspond to experimental parameters as
  described in \cite{01,07,RS}. \label{fig:sequence}}
\end{figure}
In the geometry described above, it is sufficient to consider the
one-dimensional case, where the position coordinate $x$
corresponds to the distance between the two tweezers. We will thus
assume that the motional state along the two other axis does not
change during the transport process to be described (this point is
further discussed later in this section).

The location and the depth of the minima of the
potential~(\ref{potential}) is determined by the time dependent
control parameters $A(t),B(t)$ and $d(t)$. The time evolution of
the motional degrees of freedom of a single particle in the trap
is governed by the time dependent Schr\"odinger equation
\begin{equation}
\ii \hbar \frac{d}{dt}\psi(x,t) = H(t)\psi(x,t) \label{eq:SG}
\end{equation}
with
\begin{equation}
H(t) = -\frac{\hbar^2}{2m}\frac{d^2}{dx^2} + V(x,t).
\end{equation}
In the following, distances are measured in units of a harmonic
oscillator length $a\equiv\sqrt{\hbar/m\omega}$ and energies in
units of $\varepsilon\equiv{\hbar^2/2ma^2}$. In case of
${}^{87}$Rb and $\omega=2\pi\times 100$kHz this defines a length
scale of $a = 34\,\mathrm{nm}$ and an energy scale of $\varepsilon
= 2\pi\hbar\times50\,\mathrm{kHz}$.

As in~\cite{calarco2004}, we assume that there is initially one
atom in the ground state of each well and that the barrier is
sufficiently high to prevent tunnelling between the wells. This
allows to raise rapidly the left potential well, such that at time
$t=0$ the situation depicted in figure~\ref{fig:sequence}(a) can
be created: The lowest motional state of the left atom corresponds
to the second excited state of the double well system while the
right atom is in the ground state. The moving atom can then be
adiabatically transported to the right well by lowering the left
well and the barrier simultaneously (see
figure~\ref{fig:sequence}(b)), and raising the barrier again while
the left well is further lowered leading to the final
configuration at $t=T$ shown in figure~\ref{fig:sequence}(c).
During the whole process the moving atom stays always in the
second excited instantaneous eigenstate of the system which
corresponds eventually to the first excited state of the right
well while the register atom remains in the ground state.

\begin{figure}[t!]
\begin{center}
  \includegraphics[]{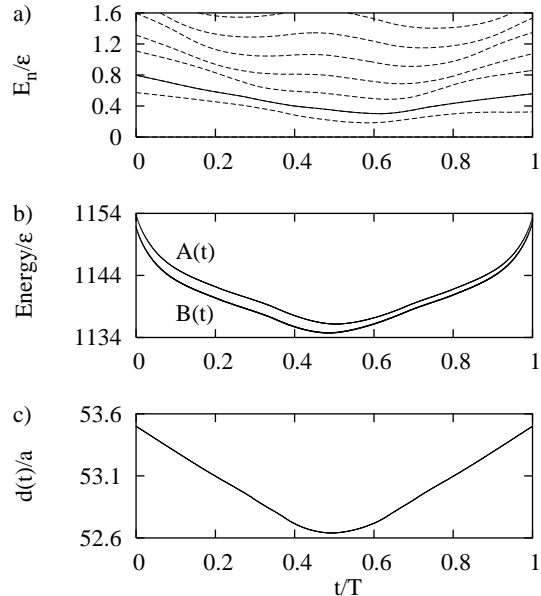}
\end{center}
\caption{%
  (a) The lowest instantaneous eigenenergies of the double-well system
  during the transport process (the ground state energy is set to
  zero). The transported atom corresponds to the solid line and the
  static atom to the zero line. (b)(c) The pulse functions $A(t)$,
  $B(t)$ and $d(t)$ control the depth and the distance of the two
  wells. In (a)-(c), the parameters for $t=0$, $t=T/2$ and $t=T$ are
  the same as in figure~\ref{fig:sequence}.
\label{fig:pulses}}
\end{figure}
The adiabatic transport is possible since by choosing appropriate
pulse functions $A(t),B(t)$ and $d(t)$, level crossings of the
eigenenergies of the system during the process can be avoided. An
example for such pulse functions, as well as the corresponding
instantaneous eigenenergies of $H(t)$, are shown in
figure~\ref{fig:pulses}. In figure~\ref{fig:pulses}(a) the time
dependent energy of the moving atom is given by the bold line (the
ground state energy is set to zero). Back transport of the moving
atom to its original position is obtained by time inversion of the
pulses.

In order to study the dynamics of the transport process we
introduce the occupation probabilities
\begin{equation}
P_n^A(t) = \left\vert \int dx\, {\psi^A}^*(x,t)
\psi_\mathrm{n}(x,t) \right\vert^2,
\end{equation}
where $\psi_\mathrm{n}(x,t)$ with $n=0,1,2,...$ is the $n$th
instantaneous eigenfunction of the double well potential. The
superscript $A\in\{M,R\}$ indicates the wavefunction of the atom
to be transported (moving atom) and the atom which is supposed to
stay located at its well (register atom), i.e. $\psi^M(x,t)$
[$\psi^R(x,t)$] is the solution of the time dependent single
particle Schr\"odinger equation (\ref{eq:SG}) with initial
condition $\psi^M(x,0)=\psi_\mathrm{2}(x,0)$
[$\psi^R(x,0)=\psi_\mathrm{0}(x,0)$] as shown in
figure~\ref{fig:sequence}(a).  The fidelities of the processes are
then given by $F^M \equiv P^M_2(T)$ and $F^R \equiv P^R_0(T)$.
\begin{figure}[t!]
\begin{center}
  \includegraphics[]{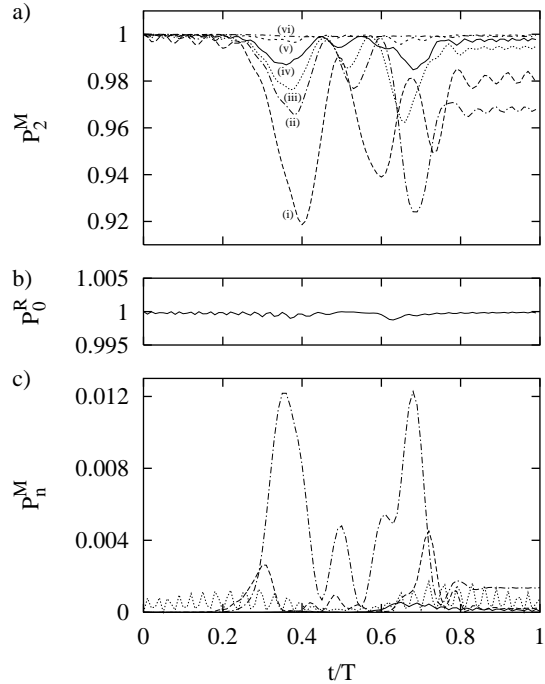}
\end{center}
\caption{
  (a) Fidelity $F^M(t)$ corresponding to the moving atom during the
  transport for $T=250\hbar/\varepsilon$ (i),
  $T=350\hbar/\varepsilon$ (ii), $T=400\hbar/\varepsilon$ (iii),
  $T=500\hbar/\varepsilon$ (iv), $T=1000\hbar/\varepsilon$ (v),
  $T=5000\hbar/\varepsilon$ (vi).  (b) Fidelity $F^R(t)$
  corresponding to the register atom during the transport for
  $T=500\hbar/\varepsilon$.  (c) Probabilities $P_0^M(t)$ (solid
  line), $P_1^M(t)$ (dash-dotted line), $P_3^M(t)$ (dashed line) and
  $P_4^M(t)$ (dotted line) of finding the moving atom in the
  respective eigenstates during its transport.
\label{fig:fidelity}}
\end{figure}
%

For the example shown in Fig.~\ref{fig:pulses} we get $F^M =
99.7\%$ for propagating the moving atom wavefunction and $F^R =
99.9\%$ for propagating the register atom wavefunction from $t=0$
to $t=T=500\hbar/\varepsilon$.  In the case of Rubidium this would
correspond to a time $T=1.6\,\mathrm{ms}$.

Figure~\ref{fig:fidelity} shows the probability $P^M_2(t)$ that
the moving atom remains in the second instantaneous eigenstate
during the transport for various operation times $T$. For
$T=500\hbar/\varepsilon$ the fidelity is always greater than
$98.7\%$ and the corresponding occupation probability $P^R_0(t)$
of the register atom (shown in figure \ref{fig:fidelity}(b)) is
always larger than $99.9\%$. The corresponding probabilities of
finding the moving atom in the ground, the first excited, the
third excited and the fourth excited instantaneous eigenstates are
displayed in figure~\ref{fig:fidelity}(c). The occupation
probabilities of higher excited states are smaller than
$4\times10^{-5}$ and are not shown. An excitation energy of
$100$kHz, which would be required for the excitation of radial
motional states, would correspond at least roughly to the eighth
excited state along the axis of motion. The occupation probability
of this (and higher) state(s) is found to be smaller than
$5\times10^{-9}$ which justifies the one-dimensional model used in
this paper.

In the following we will discuss the influence of experimental
imperfections, especially variations in the laser intensities,
which are proportional to the pulse functions $A(t)$ and $B(t)$,
and variations in the distance of the lasers creating the double
well potential, which affect the pulse function $d(t)$.  Motivated
by the experimental conditions we assume variations of the pulse
functions of the form
\begin{eqnarray}
\tilde d(t) &=& d(t) + \delta d \sin(\Omega t) \\
\tilde A(t) &=& A(t) + \delta A \sin(\Omega t) \\
\tilde B(t) &=& B(t) + \delta B \sin(\Omega t)
\end{eqnarray}
with $\Omega = 2\pi/1$ms.  Assuming a drift $\delta d = 1nm$ while
$\delta A=\delta B=0$ leads for $T=500\hbar/\varepsilon$ to a
slight reduction of the fidelity to $F^M = 99.4\%$ and a higher fidelity $F^R$. A variation of $\delta A = 0.1\varepsilon$ while $\delta B =
0$ results in $F^M = 99.5\%$ and $F^R = 99.9 \%$.

If $A(t)$ and $B(t)$ undergo the {\em same} perturbation, i.e.
$\delta A = \delta B$, the shape of the
potential~(\ref{potential}) does not change significantly if the
variation is not too large (except for an approximately constant
shift of the potential). Therefore, the level spacing as shown in
figure~\ref{fig:pulses}(a) remains roughly the same and it is
expected that the transport can be done as fast as without
fluctuations. This behavior is confirmed by numerical simulations:
Assuming $\delta A = \delta B = 10\varepsilon$, which corresponds
to variations of the laser intensity of approximately $1\%$, we
get the fidelities $F^M = 99.7\%$ and $F^R = 99.9\%$.

This analysis shows that the current transport scheme is
relatively insensitive to noise which affects both parameters,
$A(t)$ and $B(t)$, in the same way, while it is more sensitive to
different perturbations in these parameters. In this case, level
crossings in the energy diagram~\ref{fig:pulses} can appear,
leading to significant leakage into higher excited states, which
would require a more sophisticated engineering to be controlled
and will be a subject of future investigations.

\section{Quantum gates by optimal control of molecular interactions}

Performing gate operations requires a strong molecular interaction
between atoms. They can be coupled to molecular states either by
means of Feshbach resonances \cite{Feshbach} or through Raman
photo-association laser pulses \cite{photoassociation}. For the
sake of concreteness, we focus here on Feshbach resonances --
however, all of our arguments can be adapted, e.g., to Raman
photo-association. We consider $^{87}$Rb atoms.

Feshbach resonances occur when a bound molecular state $|n\rangle$
crosses the dissociation threshold for a state having the same
quantum numbers \cite{Feshbach} while changing an external
magnetic field $B$. Close to resonance, the scattering length
varies as
\begin{equation}
A(B)=A_{bg}\left( 1-\frac{\Delta_n}{B-B_0}\right) ,
\end{equation}%
where $A_{bg}$ is a non resonant background scattering length,
$B_0$ is the resonant magnetic field, and $\Delta_n$ is the width
of the resonance. The resonance energy varies almost linearly with
the field
\begin{equation}
\varepsilon_n(B)=s_n(B-B_0), \label{resenergy}
\end{equation}%
with a slope $s_n$. We are interested in the dynamics of such a
system in a confined geometry. Following \cite{Mies00}, we shall
model it by the effective Hamiltonian
\begin{equation}
H_{\rm res}=\varepsilon_n(B)|n\rangle\langle
n|+\sum_v(v\hbar\nu|v\rangle\langle v|+V_v|v\rangle\langle n|+{\rm
h.c.}),
\end{equation}
where the $|v\rangle$'s are the trapped relative-motion atomic
eigenstates of an isotropic harmonic oscillator trap having
frequency $\nu$. The couplings to the resonance are
\begin{equation}
\label{couplings} V_v=2\hbar\nu
\sqrt{\sqrt{4v+3}\;a_{bg}\delta_n/\pi}
\end{equation}
with $a_{bg}\equiv A_{bg}\sqrt{m\nu/\hbar}$,
$\delta_n\equiv\Delta_n s_n/(\hbar\nu)$. In a different geometry,
for instance in an elongated trap characterized by a ratio
$\gamma$ between the ground level spacings in the transverse and
in the longitudinal potential, the couplings can be calculated by
projection on the corresponding eigenstates \cite{calarco2004}.
Accurate values for the resonance parameters $\Delta_n$ and $B_0$,
as well as for $A_{bg}$, are now available from both theoretical
calculations and recent measurements \cite{Verhaar02}.

The possibility of controlling the resonance energy via an
external magnetic field, as described by Eq.~(\ref{resenergy}),
provides a straightforward way to steer the interaction between
the atoms. Indeed, the coupling to a specific resonant state
$|n\rangle$ is only effective for a particular entrance channel,
{\em i.e.} a specific combinations of atomic hyperfine states
(that is, of logical qubit states in our case), while in general
all other channels will be unaffected by the resonance. Thus the
resonance-induced energy shift will cause a two-particle phase to
appear only for that particular two-qubit computational basis
state.

\begin{figure}[t!]
\begin{center}
  \includegraphics[]{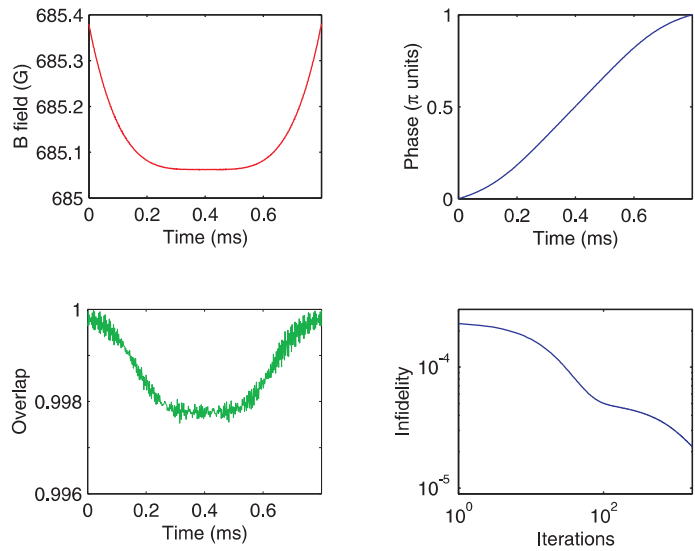}
\end{center}
\caption{Two-qubit gate operation via optimal magnetic field
  control: optimized field time dependence (top left); overlap between
  initial and evolved state (bottom left); accumulated two-particle
  phase $\varphi$ (top right); decrease of the infidelity with
  increasing iterations (bottom right).
\label{fig:gate}}
\end{figure}
We will identify our qubit logical states with the
clock-transition states
\begin{equation}
\left\vert 0 \right\rangle \equiv \left\vert
F=1,m_{F}=0\right\rangle,\quad \left\vert 1 \right\rangle \equiv
\left\vert F=2,m_{F}=0\right\rangle.
\end{equation}%
The main advantage of this choice is that the qubit states are not
sensitive to the magnetic field, and hence not subject to
decoherence due to its fluctuations. We will use the resonance for
the channel $|00\rangle$ occurring around $B_0=685$ G, having a
width $\Delta_n=16$ mG. For obtaining a two-qubit gate, the
magnetic field is ramped across $B_0$, and eventually tuned out of
the Feshbach resonance again, getting the following truth table
for the operation:
\begin{eqnarray}
\ket{00}&\rightarrow &e^{\ii \varphi}\ket{00}, \nonumber\\
\ket{01}&\rightarrow &\ket{01},  \nonumber\\
\ket{10}&\rightarrow &\ket{10}, \nonumber\\
\ket{11}&\rightarrow &\ket{11},
\end{eqnarray}
where we included the phase $\varphi$ accumulated by state
$|00\rangle$ during the ramping process due to the interaction
energy shift, whose value can be adjusted by controlling the
magnetic field. If $\varphi=\pi$, a C-phase gate between the two
atoms is obtained. Note that laser addressing of single qubits is
never required throughout the procedure.

The magnetic ramping process can be even performed
non-adiabatically, provided that all population is finally
returned to the trapped atomic ground state. This can be
accomplished via a quantum optimal control technique in analogy
with the above discussion for the transport process. The control
parameter in this case is the external magnetic field $B$. Care
has to be taken in optimizing not only the absolute value of the
overlap of the final state onto the goal state, but also its phase
$\varphi$. Fig.~\ref{fig:gate} shows the optimization results for
a trap with transverse frequencies of 100 kHz and a longitudinal
frequency of about 25 kHz, corresponding to the right well of
Fig.~\ref{fig:sequence}. The final infidelity is about
$2\times10^{-5}$ in this case.

\section{Outlook}

We have described a scheme using moving tweezers and
state-dependent controlled collisions, which is able to implement
a quantum gate between two individual atoms with a high fidelity.
The sensitivity of the scheme to intensity or position
fluctuations has been examined, and the controlled motion  is
found to be very tolerant to ``common-mode" noise between the two
tweezers. It is even relatively tolerant to differential noise,
because the overall process is close to adiabatic, and designed in
such a way as to avoid unwanted level crossings.

The magnetic field which is used here to obtain the Feshbach
resonance would be ultimately very advantageously replaced by an
optical field \cite{schlyap,grimm}, which can be switched on and
off with high speed and precision, and which will not perturb the
neighboring atoms stored in the holographic array. Though the
overall scheme is clearly not easy to implement, optimal control
techniques as used here certainly help to make it closer to
realistic.

\ack

This research was supported by a Marie Curie Intra-European Fellowship
within the 6th European Community Framework Programme,
the  RTN ``CONQUEST'',
and by the IST/FET/QIPC projects ``QGATES'' and ``ACQP''. The Institut
d'Optique group acknowledges partial support from ARDA/NSA.

\end{document}